\documentclass[aps,prb,twocolumn,superscriptaddress,preprintnumbers, amsmath,amssymb
]{revtex4-2}

\usepackage{graphicx}
\usepackage{dcolumn}
\usepackage{bm}
\usepackage[T1]{fontenc}
\usepackage{physics}
\usepackage{ulem}

\usepackage{hyperref}
\usepackage{xcolor}

\newcommand{\blue}[1]{\textcolor{blue}{#1}}

\begin{document}


\title{Magic phase transition and non-local complexity in generalized $W$ State}

\author{A. G. Catalano}
\affiliation{Institut Ru\dj er Bo\v{s}kovi\'c, Bijeni\v{c}ka cesta 54, 10000 Zagreb, Croatia}
\affiliation{Universit\'e de Strasbourg, 4 Rue Blaise Pascal, 67081 Strasbourg, France}
\affiliation{Dipartimento di Fisica e Astronomia "G. Galilei", via Marzolo 8, I-35131, Padova, Italy.}

\author{J. Odavi\'{c}}
\affiliation{Institut Ru\dj er Bo\v{s}kovi\'c, Bijeni\v{c}ka cesta 54, 10000 Zagreb, Croatia}
\affiliation{Dipartimento di Fisica Ettore Pancini, Università degli Studi di Napoli Federico II, via Cinthia, 80126 Fuorigrotta, Napoli, Italy}
\affiliation{INFN, Sezione di Napoli}

\author{G. Torre}
\affiliation{Institut Ru\dj er Bo\v{s}kovi\'c, Bijeni\v{c}ka cesta 54, 10000 Zagreb, Croatia}

\author{A. Hamma}
\affiliation{Dipartimento di Fisica Ettore Pancini, Università degli Studi di Napoli Federico II, via Cinthia, 80126 Fuorigrotta, Napoli, Italy}
\affiliation{INFN, Sezione di Napoli}

\author{F. Franchini}
\affiliation{Institut Ru\dj er Bo\v{s}kovi\'c, Bijeni\v{c}ka cesta 54, 10000 Zagreb, Croatia}

\author{S. M. Giampaolo}
\affiliation{Institut Ru\dj er Bo\v{s}kovi\'c, Bijeni\v{c}ka cesta 54, 10000 Zagreb, Croatia}

\date{\today}

\begin{abstract}
We employ the Stabilizer R\'{e}nyi Entropy (SRE) to characterize a quantum phase transition that has so far eluded any standard description and can thus now be explained in terms of the interplay between  its non-stabilizer properties and entanglement.
The transition under consideration separates a region with a unique ground state from one with a degenerate ground state manifold spanned by states with finite and opposite (intensive) momenta. 
We show that SRE has a jump at the crossing points, while the entanglement entropy remains continuous. 
Moreover, by leveraging on a Clifford circuit mapping, we connect the observed jump in SRE to that occurring between standard and generalized $W$-states  with finite momenta. 
This mapping allows us to quantify the SRE discontinuity analytically.

\end{abstract}

\maketitle

Entanglement has played an important role in impetuously developing our understanding of quantum many-body systems~\cite{Amico2008, Eisert2010}. 
However, over the years, it has become increasingly clear that entanglement alone is not able to capture every feature that differentiate quantum from classical systems~\cite{susskind, ESS}. 
The most relevant example of this is the fact that entanglement alone does not guarantee the so-called \textit{Quantum Supremacy}~\cite{Harrow2017}.
Indeed, several highly entangled quantum states can be obtained from a fully factorized state by circuits made of Clifford gates~\cite{Gottesman1998, Gottesman2004}, i.e. a series of operations that can be efficiently simulated on a classical computer. 
Quantum advantage is attained at the price of non Clifford resources and exponential increment in the difficulty of simulating a quantum circuit on classical computers~\cite{Bravyi2005, Harrow2017}. 

The resource beyond Clifford operations is colloquially known as \textit{magic}~\cite{GottesmanNJP, Liu2022} and its quantification is a formidable challenge for quantum information science~\cite{GottesmanNJP}. 
Recently, a computationally tractable family of measures of Magic for qubit systems called the \textit{Stabilizer R\'{e}nyi Entropies} (SREs)~\cite{LeonePRL} has been proposed. 
For a pure state $\ket{\psi}$ defined in a system of $L$ qubits, the SRE (of order $\alpha$) is defined as
\begin{equation}
 \mathcal{M}_2 (\ket{\psi} )=\frac{1}{1-\alpha}\log_{2}{ \left( \frac{1}{2^{L}}\sum_{\mathcal{P}} |\bra{\psi} \mathcal{P} \ket{\psi}|^{2\alpha} \right)}. \label{magicdefinition}
\end{equation}
Here $\mathcal{P}=\bigotimes_{i=1}^L P_i$ runs over all possible Pauli strings, $P_{i} \in \{ \sigma^0_{i}, \sigma_{i}^{x}, \sigma_{i}^{y}, \sigma_{i}^{z} \}$ with $\sigma_i^{0}$ represents the identity operator.
It was proven that SREs with R\'enyi index greater than or equal to 2 are monotones of magic for pure states~\cite{LL}.
Among the others, the one with $\alpha=2$ has acquired a prominent role, since in some cases it is experimentally achievable~\cite{MagicQC, Tirrito2024, Gullans2023}. 
Moreover, although eq.~\eqref{magicdefinition} implies a summation of $4^{L}$ expectation values, it allows for an efficient treatment with tensor networks~\cite{TobiasPRB, Collura, TobiasNON, TobiasClifford, Frau, Tarabunga2024}.

The stabilizer entropies provide a way for analyzing the complexity of quantum states in quantum many-body systems~\cite{HammaQuench, HammaPRA, Odavic2}. 
In gapped local systems, it was found that the SRE of ground states follows a volume law in which the slope can be determined using single-spin expectation values~\cite{HammaPRA}. 
On the contrary, this local behavior of the SRE disappears in the presence of long-range correlations that can induce topological frustration in the system~\cite{frust1, frust2, frust3, frust4}.  
In this work, we make a crucial step further, showing how it is possible to exploit the SRE to signal a phase transition associated with a mirror symmetry breaking that can be found in the topologically frustrated (TF) 1D anisotropic Heisenberg model (also known as TF XYZ model).
TF models are characterized by a delocalized excitation, and such a transition is associated with the acquisition of a non-zero lattice momentum by the excitation~\cite{Catalano}.
Being associated with a single delocalized excitation, any local or string order parameter will at most show a mesoscopic behavior that vanishes in the thermodynamic limit~\cite{frust2, Catalano}.
In \cite{Catalano2023_XY}, it was argued that such a transition is a second-order boundary QPT, but that analysis did not account for the quantization of momentum.
Moreover, we will show that even entanglement fails to highlight the development of ground states that violate the mirror symmetry. 
Therefore, to the best of our knowledge, this is the first instance of a ``pure magic transition''  in a deterministic quantum system at equilibrium, even if a shift between local and non-local magic has been recently observed in the dynamics of random quantum circuits~\cite{Gullans2023, HammaTransition, Fux2023, BeriTransition}.

The reason for this ability of the SRE is that it sums the expectation values of all possible Pauli strings.
Therefore, even if individual putative order parameters display only a mesoscopic behavior, the SRE can signal the emergence of a ground state with a non-zero lattice momentum even in the thermodynamic limit.
Following the same reasoning, it is conceivable that other combinations of strings, with different physical meanings, could detect this transition. 
The most natural example that comes to mind for our case is the momentum operator: like the SRE, when written in terms of spin operators, it is a non-local operator that is written as the sum of a super-extensive number of Pauli strings~\cite{Maric2020}.
While it would detect the TF transition in a disorder-free system, the momentum operator is not clearly definable without translational invariance, while SREs are. 
Hence, thanks to its generality, we pose that the analysis of SREs can constitute a new, general tool for the study of unusual phase transitions characterized by non-local behaviors in complex systems defined in terms of qubits or qudits~\cite{Frau2024}, in an alternative and complementary way to entanglement.

Thus, our results go well beyond TF systems, extending, for instance, to all those models characterized by a finite number of excitations, such as the ANNNI model~\cite{ANNNI}. 
In our case, to better understand the SRE, we exploit a mapping between the lowest energy states of the TF XYZ chain close to the classical point (where the Hamiltonian consists of mutually commuting terms) and generalized $W$-states in which a phase characterizes each element.
As we will show, in the thermodynamic limit, the SRE remains phase-dependent, while the entanglement and expectation values of any Pauli string are insensitive to it.

The implications of these results are two-fold: on one side, it shows that inhomogeneous $W$-states may have higher complexity (as measured by SRE) while retaining the same entanglement properties. 
On the other, generalizing the results in~\cite{Odavic2} for magic and in~\cite{frust1} and~\cite{frust3} for entanglement, we will show that for the ground state of the TF XYZ spin chains, both quantities can be written as the sum as the contribution from the corresponding non-frusterated chain and that of the generalized $W$-state. 
In this way, the jump in complexity in the latter due to a finite phase equals that of the frustrated chain, providing a further link for this elusive quantum phase transition.

\begin{table}[t]
	\centering
	\renewcommand{\arraystretch}{1.3}
		\begin{tabular}{||c|l|c|c||}
			\hline
			State & Example & Ent. & SRE \\
			\hline
			$\ket{W_0}$
			& \begin{scriptsize}
				$\begin{array}{lcl} 
					\frac{1}{\sqrt{L}} & \Big( & \ket{10000\ldots} \\ 
					& + &  \ket{01000\ldots} \\
					& + & \ket{00100\ldots} + \ldots \Big)
				\end{array}$ 
			\end{scriptsize}
			& Eq.~\eqref{SWp} & Eq.~\eqref{TDexpression} \\
			\hline
			$\ket{W_p}$
			& \begin{scriptsize}
				$\begin{array}{lcl} 
					\frac{1}{\sqrt{L}} & \Big( & e^{\imath p} \ket{10000\ldots} \\ 
					& + & e^{\imath 2 p} \ket{01000\ldots} \\
					& + & e^{\imath 3 p} \ket{00100\ldots} + \ldots \Big)
				\end{array}$ 
			\end{scriptsize} & Eq.~\eqref{SWp} & Eq.~\eqref{eq:extramagic} \\
			\hline
			$\ket{\omega_0}$
			& \begin{scriptsize}
				$\begin{array}{lcl} 
					\frac{1}{\sqrt{2L}} & \Big( & 
					\ket{\uparrow \uparrow \downarrow \uparrow \downarrow\ldots} \\ 
					& + & 
					\ket{\downarrow \downarrow \uparrow \downarrow \uparrow \ldots} \\
					& + & 
					\ket{\downarrow \uparrow \uparrow \downarrow \uparrow \ldots}  + \ldots \Big)
				\end{array}$ 
			\end{scriptsize} & Eq.~\eqref{Somegap} & Eq.~\eqref{TDexpression} \\
			\hline
			$\ket{\omega_p}$
			& \begin{scriptsize}
				$\begin{array}{lcl} 
					\frac{1}{\sqrt{2L}} & \Big( & e^{\imath p} 
					\ket{\uparrow \uparrow \downarrow \uparrow \downarrow\ldots} \\ 
					& + & e^{\imath p} 
					\ket{\downarrow \downarrow \uparrow \downarrow \uparrow \ldots} \\
					& + & e^{\imath 2 p} 
					\ket{\downarrow \uparrow \uparrow \downarrow \uparrow \ldots}  + \ldots \Big)
				\end{array}$ 
			\end{scriptsize} & Eq.~\eqref{Somegap} & Eq.~\eqref{eq:extramagic} \\
			\hline
		\end{tabular}
		\label{StateTable}
		\caption{Recap of the states used in this work. $\ket{W_0}$ and $\ket{W_p}$ are respectively the standard and generalized $W$-states defined in eq.~\eqref{eq:Wpstate}: they have the same amount of entanglement in the thermodynamic limit, but a different SRE for finite/zero $p$. $\ket{\omega_0}$ and $\ket{\omega_p}$ are linear superposition of kink states as in eq.~\eqref{eq:wkinkstate} and can be obtained from their corresponding $W$-states through the Clifford circuit in eq.~\eqref{cliffordcircuit}. Therefore, they share the same amount of magic, but not of entanglement. The ground state of the XYZ chain in eq.~\eqref{XYZ} has entanglement and magic amount of resource given by its non-frustrated (non-universal) counterpart plus the contribution form the corresponding $\ket{\omega_{0,p}}$, see eq.~\eqref{eq:Magic_sum}. }
\end{table}

As it is well known, $W$-states~\cite{Dur} play a pivotal role in quantum information~\cite{Wotters, Yeh, Gioia} 
finding applications in various quantum algorithms, such as Grover's~\cite{Grover} or quantum error correction \cite{Faist2020,Hayden2021}, quantum communication~\cite{Tsai}, and error detection~\cite{Vijayan}.
The family of generalized $W$-states (gWs) that we consider in this paper is the one that still preserves the invariance under spatial translation that reads
\begin{equation}
	\label{eq:Wpstate}
    \ket{W_{p}}=\frac{1}{\sqrt{L}}\sum_{j=1}^L e^{\imath p j} \sigma^z_j\ket{-}^{\otimes L}.
\end{equation}
Here $L$ is the number of qubits in our system, $p=2 \pi\ell/L$ with integer $\ell$ running from $-\frac{L-1}{2}$ to $\frac{L-1}{2}$ given the quantized phase, and $\ket{\pm}$ are the eigenstates of the Pauli operator $\sigma^x$ corresponding to eigenvalues $\pm1$. 
It is easy to see that the entanglement properties of the states in eq.~\eqref{eq:Wpstate} are independent of $p$.
Indeed, for any bipartition, the reduced density matrix admits only two non-vanishing eigenvalues that coincide with the probabilities that the single excitation, represented by the state $\ket{+}$, is in one or the other subset. 
Such probabilities are given by $(1 \pm x)/2$, where $x$ is the difference in the number of qubits between the two subsets normalized by $L$. Thus the 2-R\'enyi entropy at half chains is 
\begin{equation}
	S_2(W_p)  = -  \log_2   \left[ \frac{ 4 L^2+(L-1)^2- 4(L-1)L}{4L^2}\right]
	\label{SWp}
\end{equation}

On the contrary, the expression of the SRE for gWs is a function of $L$ and $p$, and we evaluated it analytically to be (see appendix)
\begin{align}
    \mathcal{M}_{2} (p,L) = - \log_2{\left( - \dfrac{11 - 12L + \frac{\sin{\left( (2 - 4L) p \right)}}{\sin{(2 p )}}}{2 L^{3}}\right)}. \label{TDexpressionFinite}
\end{align}
In the limit $p \to 0$ we reach the minimum of eq.~\eqref{TDexpressionFinite} and recover the SRE for the homogeneous $W$-state:
\begin{align}
    \mathcal{M}_{2} (0,L) = 3 \log_{2}{(L)} - \log_{2} {(7 L - 6)}. \label{TDexpression}
\end{align}
Conversely, for $\ell \neq 0$ we have that eq.~\eqref{TDexpressionFinite} becomes
\begin{align}
    \left.\mathcal{M}_{2} \left( \frac{2 \pi}{L} \ell,L \right)\right|_{\ell\neq0}  \!\!\!\!\!\!\!\!=\mathcal{M}_{2} (0,L) + \log_2{\left(  \dfrac{7 L - 6}{6 L - 6} \right)}.
 \label{eq:extramagic}   
\end{align}
Note that this expression is independent of $\ell$, as long as the latter is finite: the difference $ \Delta \mathcal{M}_{2} (L) =\mathcal{M}_{2} \left( \frac{2 \pi}{L} \ell, L \right) - \mathcal{M}_{2} (0, L)$, represents a jump in SRE, associated to the state acquiring a non-zero phase (i.e. momentum), which reduces to $\log_2(7/6)$ in the thermodynamic limit. 

Let us now introduce our many-body system, the Hamiltonian of the 1D XYZ chain with a global magnetic field along the $z$-axis reads
\begin{align}
    H_{\rm XYZ} = \sum_{n=1}^{L} \sum_{\alpha}
   J_\alpha \sigma_{n}^{\alpha} \sigma_{n+1}^{\alpha} + h \sum_{n = 1}^{L} \sigma_{n}^{z}. \label{XYZ}  
\end{align}
Here and in the following, we will always assume that $J_{x} = 1$ while $\vert J_{y} \vert,\, \vert J_{z} \vert < 1 $. 
Within Frustrated Boundary Conditions (FBCs)~\cite{Cador, Laumann, Dong, frust1, frust2, Maric2020}, the number of sites is taken to be odd ($L=2M+1$ for $M \in \mathbb{N}$) and periodic boundary conditions are assumed ($\sigma^\alpha_n=\sigma^\alpha_{n+L},\ \forall n$). 
FBCs induce a frustration of topological origin in these spin chains, with several interesting consequences on their low-energy properties, such as the introduction of unusual long-range behavior of correlations~\cite{frust2} and the substitution of a gapped system with one in which the ground state is part of a band of $2L$ states~\cite{frust4}.

\begin{figure}
	\centering
\includegraphics[width=.79\columnwidth]{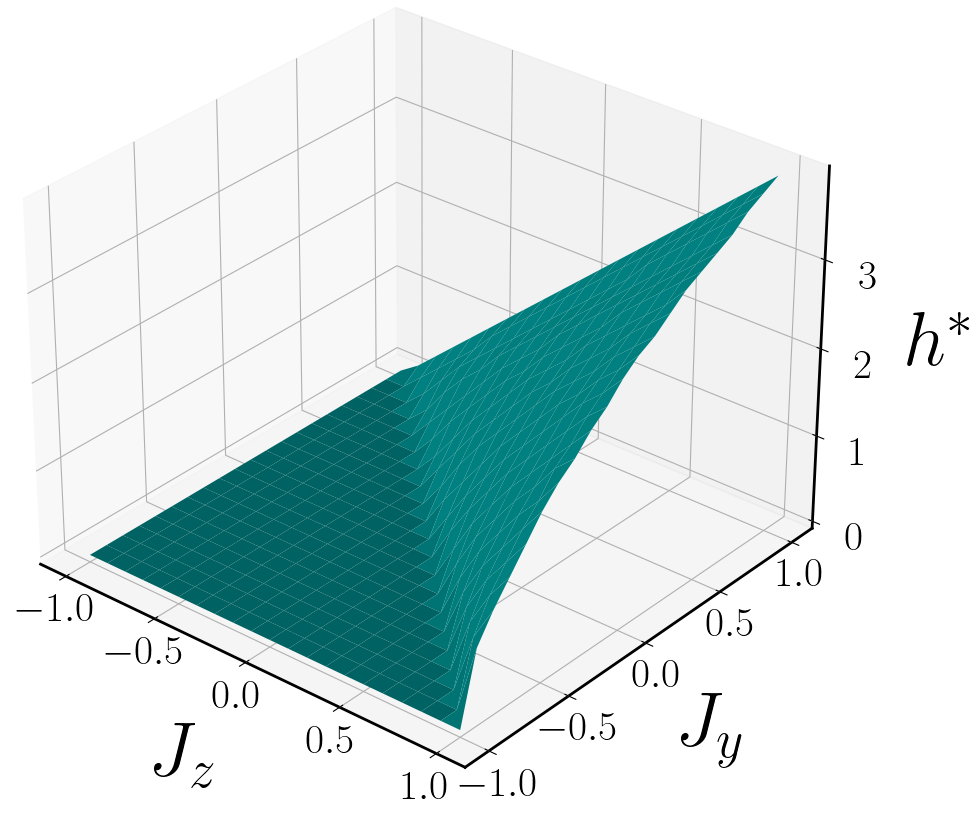}
	\caption{Value of $h^*$ as function of $J_y$ and $J_z$ ($J_x$ is assumed to be equal to 1)
	for the Hamiltonian in eq.~\eqref{XYZ}. The data is obtained numerically looking at the momentum of the ground state for a system made of $L=15$ spins. 
    For $h^{*} > 0$, choosing $\vert h  \vert < h^*$ the ground state manifold has a dimension equal to 2 and is spanned by states with finite, opposite momenta $p\neq 0$.}
   \label{fig:chiral_phase}
\end{figure}

The non-trivial response of AFM spin chains to FBC is also witnessed by an excess of bipartite entanglement beyond the area-law contribution~\cite{frust1, Odavic1, frust3}. 
While these properties characterize the whole frustrated phase, accordingly with~\cite{Catalano}, assuming $J_z\ge-J_y$ there exists a critical value of the external magnetic field $h^{*} > 0$ (see Fig.~\ref{fig:chiral_phase}) such that for $\vert h  \vert < h^*$ the ground state manifold becomes twofold degenerate and spanned by states with finite, opposite momenta $p\neq 0$.
Interestingly, the physics of the whole frustrated phase can be described in a quasi-particle picture through a single delocalized excitation in the ground state of frustrated chains.
While in most cases this excitation carries zero momentum, below $h^{*}$, where the ground-state manifold is at least two-fold degenerate, it owns a nonvanishing one.

In the beginning, let us focus on a simple case where the system can be studied analytically.
This case is represented by the system close to the classical point. The classical point is when $J_x$ is the only non-vanishing Hamiltonian parameter and eq.~\eqref{XYZ} reduces to a sum of mutually commuting terms, i.e., to a classical Hamiltonian.
Close to this point, exploiting perturbation theory, we obtain that the elements of the lowest-energy band can be written as \textit{kink states} 
\begin{align}
	\label{eq:wkinkstate}
	\ket{\omega_{p}}=\frac{1}{\sqrt{2L}}\sum_{k=1}^L e^{i p k} (\ket{k} + \ket{k'}),
\end{align}
where $p$ is the quantized momentum, i.e. $p=2 \pi \ell/L$, with $\ell=-\frac{L-1}{2},\ldots, \frac{L-1}{2}$. 
The kinks are embedded in Ne\'{e}l order states and are made of the union of two extensive sets of states defined as $\ket{k} = {\cal T}^{k} \bigotimes_{j=1}^{M} \sigma_{2j}^z \ket{-}^{\otimes L} $ and $
\ket{k'} = {\cal T}^{k} \bigotimes_{j=1}^{M} \sigma_{2j}^z \ket{+}^{\otimes L}$ with $k$ and $k'$ running from $1$ to $L$. 
Turning on $J_y$ $J_z$ and/or $h$ in the proximity of the classical point, we can have either a unique or a two-fold degenerate ground state. 
In the first case, the lowest energy is obtained by setting $p=0$, while in the other case, the two ground states display equal but opposite momenta $p$. Note that a $p=0$ state is invariant under the reflection of the system with respect to any of its sites, while a state with finite momentum breaks this mirror symmetry.

As shown in the appendix, limiting ourselves to partitions ($A|B$) composed of connected subsets, regardless of their dimensions, the reduced density matrix $\rho_A(p)=\mathrm{Tr}_B(\ket{\omega_{p}}\!\!\bra{\omega_{p}})$ admits only 4 non-zero eigenvalues:
\begin{eqnarray}
\label{eq:eigval}
 \lambda_{1,\ldots,4}  & = &  \frac{1}{4L} \Big( L+ 2 \gamma \cos( p a )  \pm \Big. \\
   & \pm  & \Big. \sqrt{(L-2 a)^2+ 4L (1+\gamma \cos( p a )) -4\sin^2( p a )  } \Big) \,,\nonumber
\end{eqnarray}
where $a=\mathrm{dim}(A)$ and $\gamma=\pm 1$. 
From eq.~\eqref{eq:eigval} it is evident that the momentum-dependent terms scale at most with the inverse square root of $L$ and thus vanish in the thermodynamic limit. 
Therefore, for diverging $L$, the entanglement does not depend on $p$.
To provide an example, setting $a=(L-1)/2$ and evaluating the 2-R\'enyi entropy of $\ket{\omega_{p}}$ we obtain
\begin{equation}
    S_2(\omega_{p})  = -  \log_2   \left[ \frac{1 + L (4 + L )  +  4  \cos(p) }{4L^2}\right]
    \label{Somegap}    
\end{equation}
that becomes independent on $p$ when $L$ diverges.

On the contrary, SRE works perfectly to detect the emergence of a ground state with a finite momentum. 
To verify this result, it is enough to observe that the states in~\eqref{eq:wkinkstate} can be obtained from the ones in~\eqref{eq:Wpstate} by applying to the second the (SRE preserving) Clifford circuit introduced in Ref.~\cite{Odavic2} 
\begin{equation}
\!\! \hat{\mathcal{S}}\!=\!\!\prod_{j=1}^{L-1} \! \mathbf{C}(L,L\!-\!j)\!\left(\prod_{j=1}^M \!\sigma_{2 j-1}^z\!\!\right)\! \mathbf{H}(L) \sigma_L^z\! \prod_{j=1}^{L-1} \! \! \mathbf{C}(j,\!j\!+\!1) \Pi^z
 \label{cliffordcircuit}
\end{equation}
Here $\text{H}(j) \equiv \frac{1}{\sqrt{2}}(\sigma^x_j+\sigma^z_j)$ is the {\it Hadamard gate} on the $j$-th qubit, while  $\text{C}(j,l) \equiv \exp\left[\imath \frac{\pi}{4}(1-\sigma^x_j)(1-\sigma^z_l)\right]$ is the {\it CNOT gate} on the $l$-th qubit controlled by the value of the $j$-th one and $\Pi^z=\bigotimes_{j=1}^L\sigma^z_j$ is the parity operator along z.
Since the states $\ket{\omega_p}$ are obtained from the $\ket{W_p}$ via a Clifford circuit, they share the same value of magic. 
Therefore, when the ground state is unique and carries zero momentum, its value of the SRE is given by eq.~\eqref{TDexpression}, but when the ground state acquires a finite momentum it increases by a quantity that stays finite even in the thermodynamic limit. 

\begin{figure}
	\centering
	\includegraphics[width=.99\columnwidth]{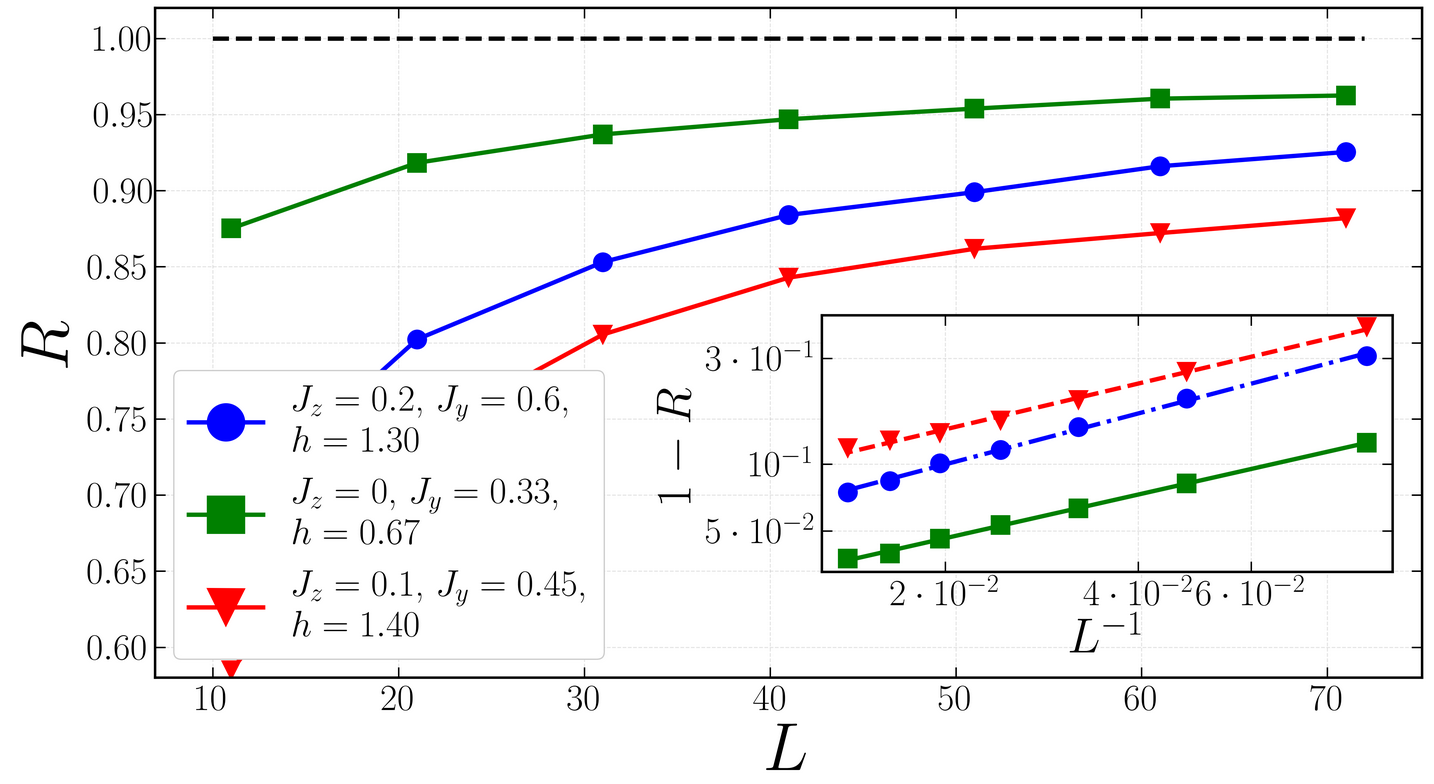}
\caption{The ratio $R(p,L)$ defined in eq.~\eqref{rfun}, as a function of $L$ for different sets of parameters. In both cases, we observe a power-law convergence of $R(p,L)\to1$ for $L\to\infty$, as highlighted in the inset plot, where we plot $1-R(p,L)$ as a function of $L^{-1}$ in log-log scale. }
	\label{fig:R_plot}
\end{figure}

To extend this result to the whole frustrated phase we show that in the thermodynamic limit it is possible to write the SRE of the ground state of a topologically frustrated spin chain $\ket{g^{TF}}$ as the sum of the SRE of the ground state of the corresponding non-frustrated model $\ket{g^{NF}}$ plus the SRE of a g$W$s, i.e. 
\begin{equation}
\label{eq:Magic_sum}
\mathcal{M}_2\big(\ket{g^{TF}}\big)=\mathcal{M}_2\big(\ket{g^{NF}}\big)+\mathcal{M}_2 \big(\ket{W_p}\big).
\end{equation}
We prove this decomposition numerically. 
To perform a meaningful finite-size scaling analysis, we need system sizes that go beyond the capabilities of exact diagonalization techniques.
Larger chains typically involve an exponential increment in the number of correlation functions but, recently, several methods have been introduced to estimate SRE using matrix product states (MPS) representations~\cite{TobiasPRB, Collura, Tarabunga2024}. 
In our case the most suited approach is the one proposed in~\cite{Tarabunga2024} since its worse scaling with the MPS bond dimension, compared for instance to~\cite{Collura}, is compensated by the possibility of avoiding any statistical sampling on the distribution of Pauli strings. The latter is problematic since it does not converge easily in the frustrated case, due to the emergence of a multi-peaked distribution for the correlation functions. 
Therefore, we first compute the chain's ground-state chain in the MPS form using a density matrix renormalization group (DMRG) algorithm~\cite{Schollwock, Orus, GottesmanArxiv} and then use it to evaluate its SRE. 
We follow this procedure both to determine the ground state of the topologically frustrated chain $\ket{g^{TF}}$, and the one of the corresponding non-frustrated model $\ket{g^{NF}}$, obtained by inverting the signs of $J_x$ and $J_y$. 

\begin{figure}
	\centering
\includegraphics[width=.99\columnwidth]{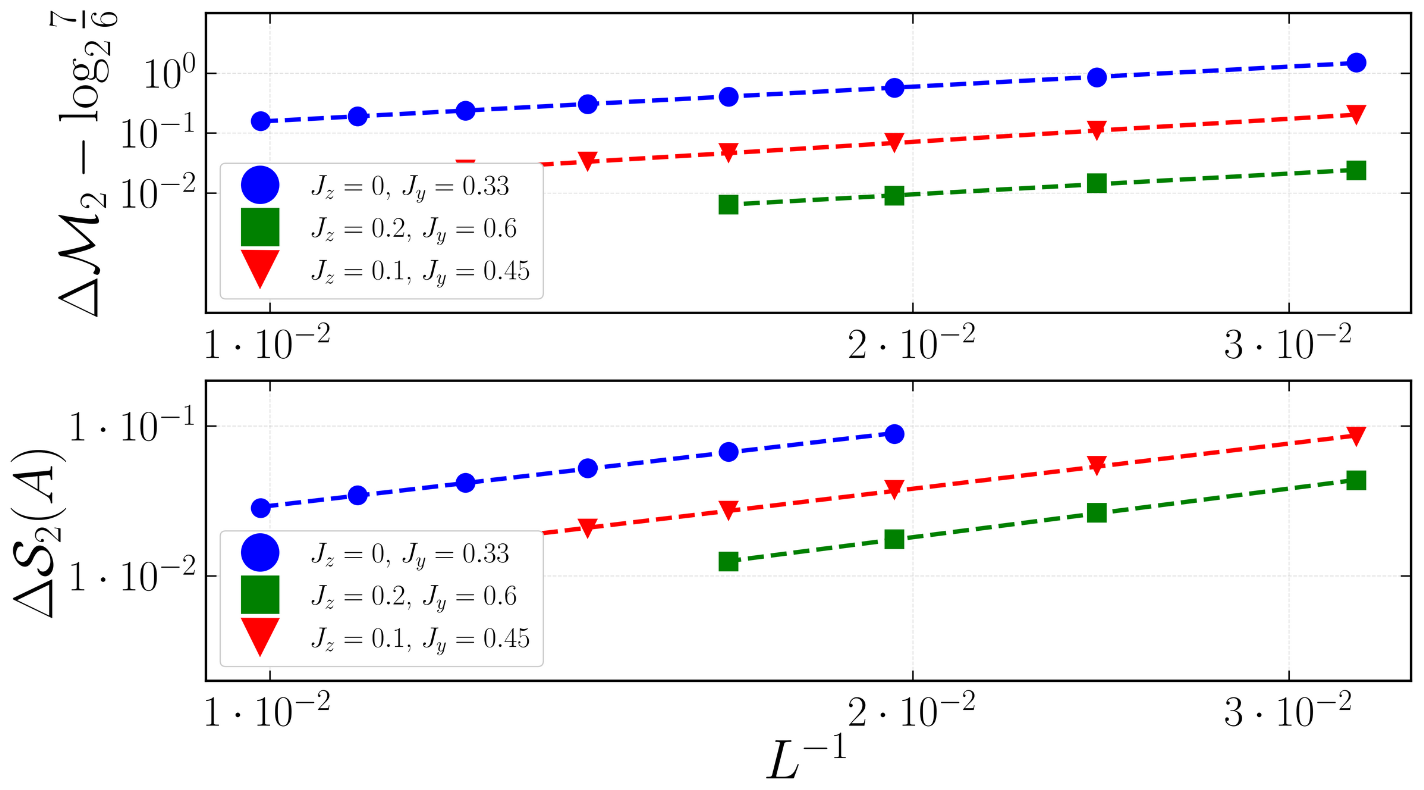}
	\caption{Finite-size scaling analysis of the discontinuity in the SRE (top) and in the entanglement (bottom) for different sets of anisotropies ($J_x=1$ in all analyted cases ). Both the two quantities show a power-law decay to the thermodynamic values that are, respectively $\log_2(7/6)$ and 0. The entanglement is evaluated with the 2-R\'{e}nyi entropy and the data plotted are associated at the partition $(A|B)$ in which $A$ is a connected subsystem made of $(L-1)/2$ spins.}
	\label{fig:magic_jump}
\end{figure}

In Fig.~\ref{fig:R_plot} we plot the quantity
\begin{equation}
R(p,L)=\frac{\mathcal{M}_2\big(\ket{\psi^{TF}}\big)}{\mathcal{M}_2\big(\ket{\psi^{NF}}\big)+\mathcal{M}_2g(\ket{W_p}\big)}, \label{rfun}
\end{equation}
that clearly approaches unity as $L \to \infty$ hence proving  eq.~\eqref{eq:Magic_sum}.
The data from this analysis would already be sufficient to prove that the phase transition associated with the violation of the mirror symmetry is highlighted by a finite gap of the magic. 
However, it is also interesting to provide direct verification.
Therefore, we performed a finite-size scaling analysis of the jump in magic at the transition point $h^*$ for several sets of parameters. 
To realize this analysis, we fix the values of the anisotropies, determine numerically $h^*$ and plotted the difference in the SRE soon after and soon before this point. 
In all analyzed cases, the numerical data show a power-law convergence of the amplitude of the discontinuity to the analytically computed value of $\log_2(7/6)$, as shown in Fig.~\ref{fig:magic_jump}. 
In contrast, the discontinuity in the bipartite entanglement shows a power-law convergence to 0, implying that, in the thermodynamic limit, it is unable to highlight the presence of the phase transition.
This confirms that the SRE witnesses the quantum phase transition associated with the violation of the mirror symmetry in topologically frustrated spin chains, which could therefore be classified as a \textit{magic (SRE) transition.} 

Let us stress that no observable that can be written as a product of Pauli operators could detect this phase transition. In fact, in the appendix we prove, for the gWs $\ket{W_{p}}$ in eq. \eqref{eq:Wpstate}, that if the expectation value of a string depends on the phase $p$, then it is mesoscopic and vanishes in the thermodynamic limit. Since the Clifford circuit that transforms states $\ket{W_{p}}$ into $\ket{\omega_{p}}$, also maps a Pauli string into another Pauli string, only operators involving an extensive amount of Pauli strings can detect a finite momentum $p\neq0$ also in the many-body model.

Summarizing, we introduced a generalization of $W$-states that promotes a finite momentum. While preserving the value of entanglement of the $W$-states, they possess a greater degree of complexity as highlighted by the SRE.
This generalization of $W$-states is extremely relevant in topologically frustrated 1D systems since they can be mapped, through a Clifford circuit into the elements of the lowest energy band close to the classical point. 
Then, we showed that, since the complexity of the ground states of topologically frustrated chains can be decomposed as the sum of a non-frustrated component and that of the g$W$s, the transition separating the region of zero momentum ground state from that with finite momenta can be characterized by a discontinuity in SRE. 
While it was already shown in other works that the SRE can detect measurement-induced phase transitions that are not signaled by the entanglement entropy in quantum circuits, to the best of our knowledge, our result constitutes the first instance of a quantum phase transition that can only be witnessed by the SRE in a deterministic system. 
The reason behind the fact that only the SRE can capture this quantum phase transition is probably related to the fact that the corrections induced by topological frustration on quantities like the correlation functions typically decay at least as $L^{-1}$, hence vanishing in the thermodynamic limit. 
The SRE, however, involves the sum of the expectation values of an exponential number of correlation functions and hence can display a finite jump even in the thermodynamic limit. 
It is important to stress once more that, while the Clifford mapping does not preserve the bipartite entanglement entropy and thus those of g$W$s and the spin ground states differ, they do not show discontinuities when a finite momentum appears.

The nature of this transition, being induced by boundary conditions, has remained controversial so far: the results of this work not only show the first instance of a discontinuity in SRE not accompanied by a similar one in the entanglement in a deterministic model, but further establish complexity in condensed matter/statistical physics systems as a detector of unconventional quantum phase transition. 
Of course, additional instances of such phenomenology are needed to establish whether complexity is just a proxy to detect transitions (like the entanglement entropy) or if it truly captures something fundamental, like topological order parameters.

\paragraph{Acknowledgments.---} This work was also supported
by the PNRR MUR Project No. PE0000023-NQSTI (J.O. and  A.H).  A.H. acknowledges financial
support PNRR MUR Project No. CN 00000013-ICSC. 
AGC acknowledges support from the MOQS ITN programme, a European Union’s Horizon 2020 research and innovation program under the Marie Sk\l{}odowska-Curie grant agreement number 955479. 

\appendix

\section{Entanglement Entropies for the $\omega_p$ states for a generic bipartition made of convex subsystems} \label{appendix:analytics}

As we have seen in the main text, near the classical point the ground state of a topological frustrated system can be well-approximated by a state $\omega_p$ that can be written in the form
\begin{align}
	\label{eq:wkinkstate1}
	\ket{\omega_{p}}=\frac{1}{\sqrt{2L}}\sum_{k=1}^L e^{i p k} (\ket{k} + \ket{k'}),
\end{align}
Here $p$ is the quantized momentum, i.e. $p=2 \pi \ell/L$, with $\ell=0,\ldots,L-1$ and $L$ being the (odd) length of the chain. 
The kinks are embedded in Ne\'{e}l order states and are made of the union of two extensive sets of states defined as 
\begin{eqnarray}
	\label{kinkdef}
	\ket{k^+} &=& T^{k-1}\bigotimes_{j=1}^{M}\sigma_{2j}^z\ket{+}^{\otimes L} \nonumber \\
	\ket{k^-} &=& T^{k-1}\bigotimes_{j=1}^{M}\sigma_{2j}^z\ket{-}^{\otimes L}\,.
\end{eqnarray}
In \eqref{kinkdef} $\ket{\pm}$ denote the eigenstates of $\sigma^x$ associated respectively to the positive/negative eigenvalue in the $x$-direction, $M=(L-1)/2$, while $T$ stands for the translation operator that shifts the state of the system by one single site towards the right.
For $k=1$ the ferromagnetic defect is placed between sites $1$ and $L$ while with $k>1$ the translation operator moves it around the whole chain.

Let us now consider a partition of the system $A|B$ in which both $A$ and $B$ are convex sets, i.e. ensembles of contiguous spins. 
From~\eqref{eq:wkinkstate1} we may recover the reduced density matrix obtained by projecting $\omega_p$ into $A$. 
In the quite general case $a=dim(A)\ge2$, the reduced density matrix can be written as follows.

\begin{eqnarray}
	\label{single_partition}
\rho_A & = & \mathrm{Tr}_B(\ket{\omega_p}\!\!\bra{\omega_p}) \\
 & = &\frac{1}{2 L}\left(
	\begin{array}{cccc}
		\mathbf{Q}^{(b)} & \mathbf{0}^{(b)} & V^{(b)}& W^{(b)} \\
		\mathbf{0}^{(b)} & \mathbf{Q}^{(b)} & W^{(b)}& V^{(b)} \\
		(V^{(b)})^\dagger &  (W^{(b)})^\dagger & L-b & 2 \cos(a p)  \\
		(W^{(b)})^\dagger &  (V^{(b)})^\dagger & 2 \cos(a p) & L-b 
	\end{array}
	\right) \nonumber
\end{eqnarray}

In eq.~\eqref{single_partition} $\rho_A$ is a $2a\times2a$ square matrix, and $b=a-1$. 
The reduced density matrix $\rho_A$ is not block diagonal but has a block structure and each one of these blocks has a quite regular structure.
To begin, the matrices $\mathbf{0}^{(b)}$ and $\mathbf{Q}^{(b)}$ are both $b\times b$ square matrices. 
All the elements of the first are zeros, i.e.  $\mathbf{0}^{(b)}_{m,n}=0 \, \forall m,\, n $, while the elements of $\mathbf{Q}^{(b)}$ obey to the following law $\mathbf{Q}^{(b)}_{m,n}=\exp(-\imath(m-n)p)$.
On the contrary, both $V^{(b)}$ and $W^{(b)}$ are column vectors made of $b$ rows and one single column.
The $n$-th element of $V^{(b)}$ can be written as 
$V^{(b)}_n=\exp(-\imath(b+1-n)p)$, while for $W^{(b)}$ we have $W^{(b)}_n=\exp(-\imath(L-n)p)$ 

Diagonalizing this matrix with the help of Mathematica, and testing the results so obtained with a purely numerical code, we have found that all the eigenvalues are zero except four.
These four non-vanishing eigenvalues can be put in the form
\begin{eqnarray}
	\label{eq:eigval1}
	\lambda_{1,\ldots,4} & = & \frac{1}{4L} \Big( L+ 2 \gamma \cos( p a )  \pm \Big.  \\ 
   & & \pm \Big. \sqrt{(L-2 a)^2+ 4L (1+\gamma \cos( p a )) -4\sin^2( p a)  } \Big) , \nonumber
\end{eqnarray}
where $\gamma$ is a dicotomic real parameter of modulus 1, i.e. $\gamma=\pm 1$.
The four eigenvalues are recovered considering all the possible combinations of the $\pm$ sign in front of the square root and the values of $\gamma$.
From this expression, all the different entropic measures of the entanglement can be easily recovered. 
A case in which the analytical expression of the entanglement entropy becomes very easy is the 2-R\'enyi entropy that, after some steps, can be reduced to
\begin{equation}
\!\!\!\!\!\!	S_2(a,p) \! = \!-  \log_2 \! \left[ \frac{L ( 2 + L ) \!-  \! 2 a (L - a) \!+\! 2 \cos(pa) }{2L^2}\right],
\end{equation}
from which, setting $a=(L-1)/2$, we can recover the result presented in the main text

\section{Analytic derivations of Stabilizer R\'enyi entropy for a $W_P$ state}

The main result shown in the main part of the paper is based on the behavior of the SRE on the family of $W_P$ states, which is a family that generalizes the well-known $W$ state.
To evaluate the SRE, let us start by recalling its expression for a generic pure state $\ket{\psi}$. 
It reads
\begin{equation}
	\mathcal{M}_2 (\ket{\psi} )=-\log_{2}{ \left( \frac{1}{2^{L}}\sum_{\mathcal{P}} \bra{\psi} \mathcal{P} \ket{\psi}^4 \right)}, \label{magicdefinition1}
\end{equation}
where the sum runs over all possible Pauli strings $\mathcal{P}$ that can be defined on the system.
Taking into account that any $\ket{W_p}$ state can be written as
\begin{equation}\label{eq:Wstate_app}
	\ket{W_p}=\frac{1}{\sqrt{L}}\sum_{j=1}^L e^{\imath p j}\sigma^z_j\ket{ -}^{\otimes L},
\end{equation}
we immediately see that to determine the SRE on $\ket{W_p}$, we have to evaluate the terms 
\begin{equation}
	O(\mathcal{P})=\sum_{j,k=1}^L \exp[\imath(j-k)p] \bra{-}^{\otimes L} \sigma_k^z \mathcal{P} \sigma_j^z \ket{-}^{\otimes L}
\end{equation}
It is easy to see that in the large majority of cases $O(\mathcal{P})=0$, but with two important exceptions.
The first exception is when the Pauli string is made only by the identity and $\sigma^x$ operators, i.e. when $\mathcal{P}$ becomes  $\mathcal{P}'=\bigotimes_{k=1}^L \sigma_k^{\alpha}$, where $\alpha \in \{ 0, x \}$.
In this case, the absolute value of each $O_{j,k}(\mathcal{P}')$ depends on the number $l=0,\ldots,L$ of $\sigma_k^x$ operators in the string, and it is equal to  $\|\frac{L-2l}{L}\|$.
Taking into account all the possible combinations of identity and $\sigma^x$ operators, the contribution of these terms becomes
\begin{equation}
	\label{marco_step_3}
	\sum_{\mathcal{P}'}O(\mathcal{P}')=\sum_{l=0}^{L}\left(\frac{L-2l}{L}\right)^4 \frac{L!}{l! (L-l)!}.
\end{equation} 
The second exception is represented by the Pauli strings with only two operators in the set $\{\sigma^y,\, \sigma^z\}$. 
Within this hypothesis, the Pauli string can be written as $\mathcal{P}''= \bigotimes_{k=1, k\neq i,j}^L \sigma_k^{\alpha} \otimes ( \sigma^\beta_j \sigma^\gamma_k )$ where $\alpha=0,x$ while \mbox{$\beta,\, \gamma=y,z$}.
This contribution comes from the fact that such strings are able to shift the $\ket{+}$ from the site $j$ to the site $k$ and vice versa.
When $\beta=\gamma$ both these two terms have the same sign, so giving a contribution proportional to $\cos{[(j-k)p]}/2L$.
On the contrary, when $\beta \neq \gamma$ they show opposite signs, so contributing proportional to $\sin[(j-k)p]/2L$. 
Naming $r=j-k$ and summing over all possible Pauli strings we have
\begin{eqnarray}
	\label{marco_step_4}
\sum_{\mathcal{P}''}O(\mathcal{P}'') & = & L \sum_{l=0}^{L-2}
	\sum_{r=1}^{L-1} \frac{(L-2)!}{l! (L-2-l)!} \\
    & \times & \left[ \left(\frac{2 \cos(p r)}{L}\right)^4+\left(\frac{2 \sin(p r)}{L}\right)^4\right]. \nonumber
\end{eqnarray}
Putting the two non-vanishing contributions together in the definition of the SRE of order 2, we recover, after some steps, the following expression
\begin{equation}
    \mathcal{M}_{2} (W_p) = - \log_2{\left( - \dfrac{11 - 12L + \frac{\sin{\left( (2 - 4L) p \right)}}{\sin{(2 p )}}}{2 L^{3}}\right)}. \label{TDexpressionFinite1}
\end{equation}

\section{Incompatible symmetries}

\begin{figure}
	\centering
 \includegraphics[width=0.95\columnwidth]{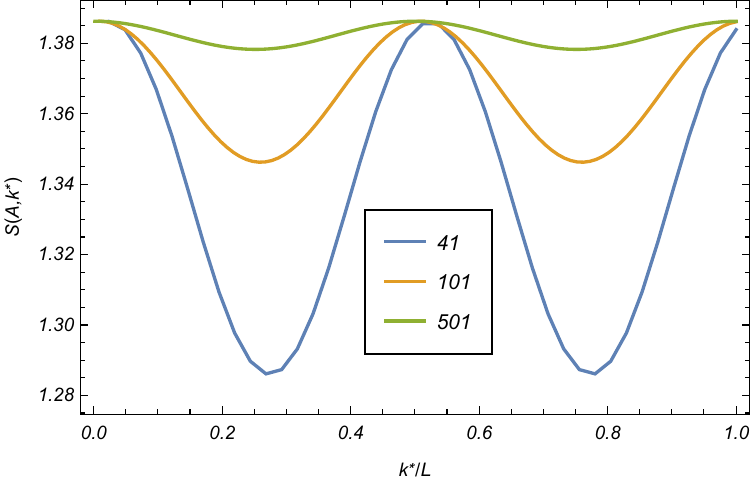}
    \hspace{.02\columnwidth}
 \includegraphics[width=.95\columnwidth]{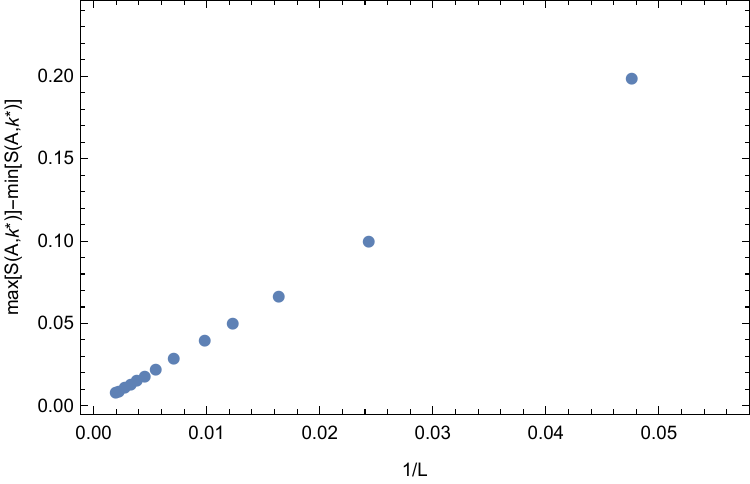}
	\caption{
    Left Panel: Dependence on the position of the entanglement properties for the states $\ket{\phi(p,\theta)}$ in \eqref{state_breaking_invariance}.
    The entanglement is evaluated considering a partition (A|B) where $A$ is made by $(L-1)/2$ contiguous spins of which the first is $k^*$,   
    $p=2\pi/L$ and $\theta=(L-1)/4$.
    Left Panel: Dependence of the amplitude of the oscillations of the entanglement properties on the dimension of the systems. 
    The difference between the maximum and the minimum of $S(A,k^*)$ is plotted as a function of $1/L$ to highlight the fact that in the thermodynamic limit it vanishes.}
	\label{fig:ent_func_position}
\end{figure}

In our paper, we considered the breaking of the mirror symmetry and observed how the violation of this symmetry, associated with the emergence of finite-momentum ground states, is evidenced by a finite jump in the value of the SRE. 
At the same time, we also observed how entanglement does not show any response to this phenomenon and, therefore, is not a quantity usable to detect the breaking of this symmetry and the quantum phase transition it entails. 
However, the phenomenology connected to this phase transition is very complex and has other aspects to highlight.

In fact, among others, the XYZ chain has two symmetries that are usually considered as trivial, namely the translational symmetry and the mirror symmetry that exchanges sites at the same distance from a reference point. In general, these two symmetries cannot be diagonalized simultaneously, and only states with zero or $\pi$ momentum can be eigenstates of both. 
However, the $\pi$ momentum is incompatible with the constraint of an odd number of spins.

Crossing the phase transition at $h=h^*$ we have a sudden change in ground state degeneracy connected with the appearance of ground states with finite and opposite momenta. In a way, the peculiarity of this phase transition, which renders it different from the usual ones, is that, after crossing it, it is not possible to select out of the degenerate ground state manifold a state that respects all the symmetries of the Hamiltonian.

\begin{figure}
	\centering
 \includegraphics[width=.95\columnwidth]{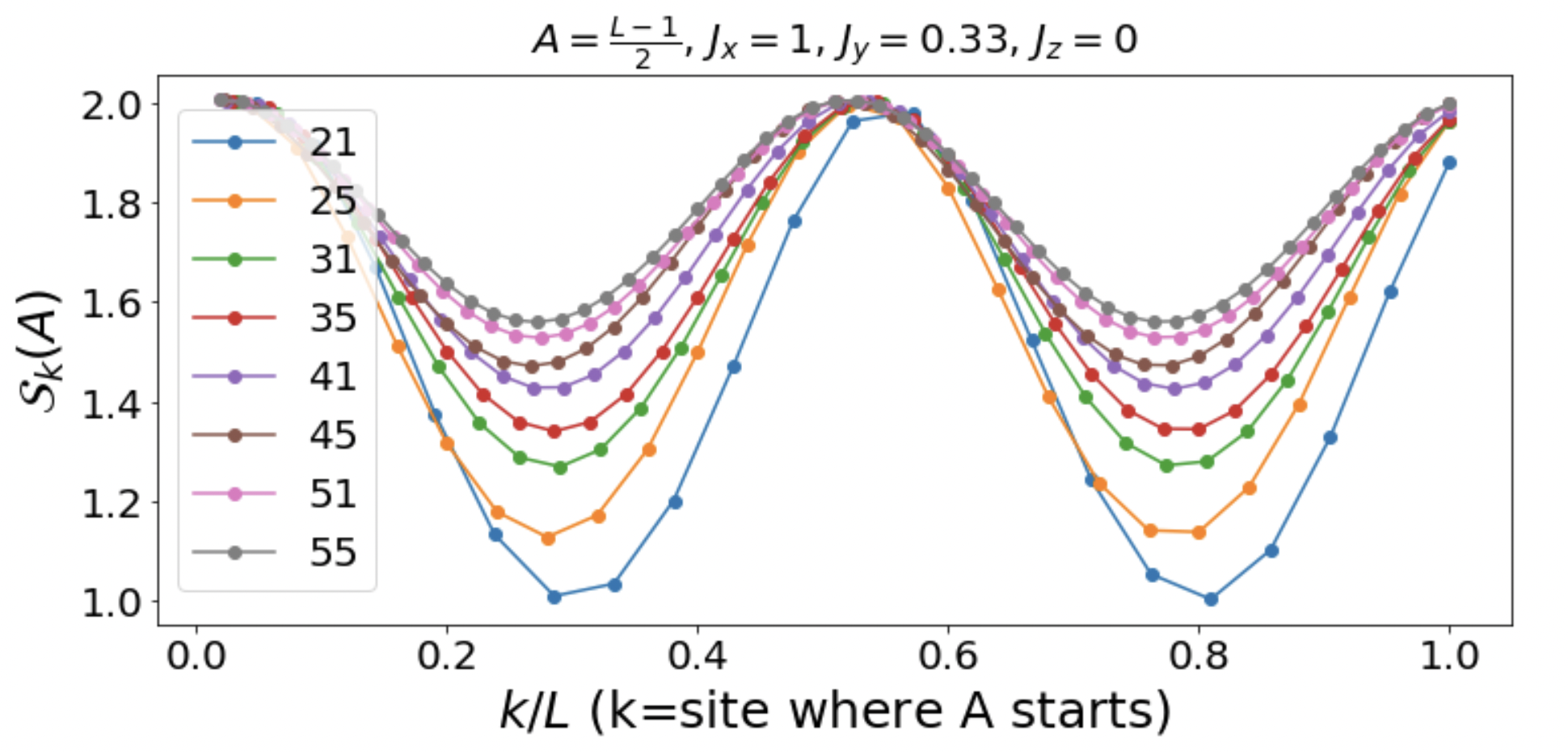}
    \caption{Dependence on the position of the entanglement properties for the ground states that preserve the mirror symmetry for different system lengths.
    The ground states considered in this plot are obtained by the DMRG algorithm setting $J_x=1$ and $J_y=0.33$. With these parameters, the ground states analyzed are the symmetric linear combination of the two ground states with momentum equal to $p=2\pi/L$. The entanglement is obtained considering a partition (A|B) where $A$ is made by $(L-1)/2$ contiguous spins of which the first is $k^*$.}
	\label{fig:ent_func_position_2}
\end{figure}

In the main part of this paper, we consider eigenstates of the momentum operator that thus break mirror symmetry: let us now look at eigenstates of the mirror symmetry, constructed as the symmetric superposition of states with opposite momenta:
\begin{eqnarray}
\label{state_breaking_invariance}
 \ket{\phi(p,\theta)} & = &\frac{1}{\sqrt{2}}(e^{-\imath \theta}\ket{\omega(p)}+e^{\imath \theta}\ket{\omega(-p)}) \\
 & = &\frac{1}{\sqrt{L}}\sum_{k=1}^L \cos(pk-\theta)(\ket{k} + \ket{k'}) \nonumber
\end{eqnarray}
It is easy to see that $\ket{\phi(p,\theta)}$ does not violate the mirror symmetry with respect to the site $\Tilde{k}=\theta/p$, but it is not an eigenstate of the spatial translation invariance symmetry of the Hamiltonian, as can be easily seen from the explicit dependence on the modulus $\cos(pk-\theta)$ from the site index $k$.

As we can appreciate from the plot in fig.~\ref{fig:ent_func_position}, the entanglement properties of the state $\ket{\phi(p,\theta)}$ depend on the value $p$ but this dependence vanishes while the dimension of the system diverges.
Indeed, in the first panel of fig.~\ref{fig:ent_func_position} we illustrate the dependence of the Von Neumann entropy for a convex bipartition $(A|B)$, where $A$  consists of $ (L-1)/2$ contiguous spins, on $k^*$ that represents the first spin of the chain in $A$.
The analysis is performed for chains of varying lengths, assuming $p=2\pi/L$ and $\theta =(L-1)/4$.
As the size of the system increases, the amplitude of the oscillations in the entanglement value decreases, vanishing in the thermodynamic limit, as can be appreciated in the second panel of the same figure. 
A numerical analysis of the data shows how, for $p=2\pi/L$, in the limit of large $L$, this amplitude can be approximated by $const/L$ where $const \simeq 4.17$.
This picture does not change significantly moving away from this perturbative regime: a similar behavior can be be seen in fig.~\ref{fig:ent_func_position_2} for the entanglement immediately below the transition point for the ground state of the spin chain obtained from DMRG. 

Finally, we should mention that the XYZ chain also possesses a $\mathbb{Z}_2$ symmetry, related to the spin flip symmetry, and represented by the parity of the magnetization along the $z$-axis. For $h > h^*$ the ground state has a fixed parity and is separated from a state with opposite parity by a gap closing polynomially with the system size~\cite{frust2} (this should be contrasted with the case without topological frustration, where the gap closes exponentially~\cite{Franchini17}).  Below the phase transition ($h < h^*$), for finite chains the lowest energy states can have different parities, with frequent level crossings between them as the Hamiltonian parameters are changed (the first of which always being at $h = h^*$) and for different chain lengths $L$. Overall, the gap between opposite parities' lowest energy states closes exponentially~\cite{Catalano}, indicating that in the thermodynamic limit, the ground state manifold doubles its degeneracy. Hence, in addition to ground states breaking the mirror and/or translation symmetry, one could also select a state with indefinite parity to study the phase transition properties. Such a choice introduces further complications and subtleties in the analysis, since the momenta of states with different parities are shifted by $\pi$ and thus several inequivalent scenarios arise, depending on the ground state choice. We believe that these subtleties are peculiar to the model under consideration and will be the subject of a separate work with detailed considerations on different finite-size scenarios, while the SRE discontinuity we reported in the main part of this paper is a general phenomenon that can help in detecting novel types of phase transitions.

\end{document}